\documentclass{article}
\usepackage{graphicx}
\usepackage{amssymb}
\usepackage{color}
\usepackage[latin1]{inputenc}
\usepackage{color}

\newcommand{\ao}{{Appl Opt \/}}
\newcommand{\oc}{ { Opt Commun \/} }

\newcommand{\figsjpg}[1]{}

\title{Adaptive aperture defocused digital speckle photography}
\author{Jose M. Diazdelacruz\\ e-mail: jmdiaz@etsii.upm.es\\Department of Applied Physics,\\ Faculty for Industrial Engineering, \\ Polytechnic University of Madrid. \\  Jose Gutierrez Abascal 2. 28006 Madrid. Spain}

\begin{document}
\maketitle

\begin{abstract}
Speckle photography can be used to monitor deformations of solid surfaces. The measuring characteristics, such as range or lateral resolution depend heavily on the optical recording and illumination set-up. This paper shows how, by the addition of two suitably perforated masks, the optical aperture of the system may vary from point to point, accordingly adapting the range and resolution to local requirements. Furthermore, by illuminating narrow areas, speckle size can be chosen independently from the optical aperture, thus lifting an important constraint on its choice. The new technique in described within the framework of digital defocused speckle photography under normal collimated illumination. Mutually limiting relations between range of measurement and spatial frequency resolution turn up both locally and when the whole surface under study is considered. They are deduced and discussed in detail.  
\end{abstract}



\section{Introduction}

When a solid object undergoes a load change, its deformation field exhibits different behaviours over the surface. For instance, when a vertical force is applied at the free end of a horizontal cantilever beam, slope changes are bigger and more uniform in the vicinity of the load than near the supported end.\cite{belendez} In other words, the ranges for spatial frequencies and magnitudes of the slope variation fields are not evenly distributed over the object surface.  

On the other hand, when optical methods are used to measure the surface deformation, they are generally tuned to provide adequate characteristics for the whole surface, so that there are one range of measurement and one lateral resolution.  This may lead to compromise solutions where the high values expected in some areas reduce the lateral resolution available even in points with anticipated lower values. 

Defocused speckle photography has long been used to measure the distribution of out-of-plane rotations over a surface under load changes. However, its measuring ranges for rotation and spatial frequency are the same over the whole area under study. Moreover, when digital recording systems are used, these values are strongly conditioned by the camera resolution. This paper describes a new enhancement of the system that makes its capabilities more adaptable both by allowing different measuring characteristics over the surface and by untying their relation to the camera resolution.  
  
  When a visible laser beam is scattered by a rough surface, the reflected light intensity exhibits a grainy distribution, called speckle pattern. The origin of this phenomenon is the interference of the light coming from all the points of the surface. If non-coherent light is used, the interference patterns vary so quickly that only the average intensity is observed and therefore speckles only appear under coherent illumination. 

The intensity pattern of the light scatterd from a rough surface can be collected by an optical system and recorded on a plane. Each set-up determines the way in which the light from different points in the surface interfere and thus the amplitudes, spatial frequencies, average speckle size and other characteristics of the intensity pattern at the recording plane.\cite{goodman} For a given set-up, the speckle distribution represents a unique signature of the surface under observation. When the surface undergoes a mechanical transformation the interference combinations at every point in the recording plane are altered, but sometimes they can be partially reconstructed in a different point at the recording plane. In this case, the speckle pattern is said to be shifted (displaced) and decorrelated (slightly modified). 

Speckle photographic techniques explore the possibilities of determining the object transformation from the speckle shift that takes place at the recording plane. When it is at the back focal plane of the optical set-up, the method is called Defocused Speckle Photography (DSP). If a digital detector (as a CCD camera) is used to record the speckle pattern then the system is said to be a Defocused Digital Speckle Photographic system (DDSP). Under normal collimated illumination, this tecnique is sensitive to out-of-plane rotations (or tilts) of the surface under observation. The speckles are displaced in the back focal plane of the lens by a distance which is proportional to the rotation angle (provided it is small).\cite{jones} It will be assumed that the investigated area is part of a rough planar surface.  

The recorded speckle pattern (or specklegram) is stored in a computer. Then a load change is applied to the object and a second specklegram is obtained. Once the two specklegrams are available, computer algorithms are applied to find the speckle displacement with sub-pixel accuracy. Finally, the distribution of the tilt throughout the area is evaluated.

When digital recording is used, speckle size considerations may play an important role in the design of the optical system. Speckles should not be smaller than the detector cell, because the speckle pattern would be spatially averaged in the recorded specklegram. This leads to a reduction in the speckle contrast and eventualy to a total loss of the pattern. On the other hand, if speckles are much wider than the detector cells, many pixels are necessary to compute the speckle displacement, leading to poorer lateral resolution. Moreover, random errors in the results of digital processing depend on speckle size, so that it should be kept as small as posible, yet taking into account the previous considerations. Some studies have been published in order to set the optimal speckle size in digital speckle photography, so that its optimum value $s^*$ lies close to 
\begin{equation} 
s^* \approx 2w
\end{equation}  
where $w$ is the pitch of the cells in the sensor array.\cite{sjo94,sjo97,sjo98} 

Speckle size in the recording plane of a defocused speckle photographic system is determined by the narrowest aperture $a_{m}$ of the light arriving at a point of the detector. 
\begin{equation} 
s = \frac{1.22 \lambda f}{a_{m}}
\end{equation} 
and therefore, it should be assured that
\begin{equation} 
a_m \approx \frac{1.22 \lambda f}{s^*}
\end{equation} 
  
As alredy stated, defocused speckle photography is used to assess the distribution of slope variations over a surface when a load change is produced. The aperture of the optical system determines both the range and the lateral resolution of measurement. It will be shown that big apertures allow larger ranges and narrower resolutions and small apertures work in the opposite way. 

The core of the method described in this paper is the use of a pair of coupled masks to illuminate and collect light from the surface into the digital camera. The masks should be suitably perforated so as to provide the speckle size, effective aperture, lateral resolution and measuring range from a distributed system approach. In this paper, the mathematical relations satisfied by the main parameters and their mutual limitations are analysed.

Previously described implementations of the method exhibit two limiting characteristics:

\begin{itemize}
\item[a)] the optical system has one aperture, and thus, the measuring range and lateral resolution are shared by all the points in the surface under observation.
\item[b)] the aperture of the system sets the already mentioned measuring characteristics and the speckle size, so that it is not possible to tune them independently.
\end{itemize}

The system presented here relays on a slightly modified implementation that improves on the afore mentioned problems. First, the optical set-up is described and then its main features are analysed. 

In short, as alredy stated, lateral resolution and maximum range depend on the effective aperture of the optical system, so that, in order to relieve the aperture determination from speckle size considerations, we use an illumination mask that produces an illumination pattern made of a discrete set of narrow circles of diameter $a$, so that the speckle size is primarily determined by $a$. Besides, a second mask is added in order to assign a different entrance pupil for every illuminated area. Therefore, it is possible to have  different speckle sizes, lateral resolutions and measuring ranges over the illuminated area. 

Recording areas and their individual cells are often rectangular, although in this work, for the purpose of simplicity, they are assumed to be square. In the following sections, the sides of the sensing area and the individual cell squares will be supposed to be $b,w$, respectively. Besides, a focal length $f$, an object to lens optical distance $d$, an  $L \times L$ square observed area and a lens aperture diameter $D$ are assumed. Further, in order to use the maximum recording area, the following relation
\begin{equation}\label{eblfd}
\frac b f =\frac L d
\end{equation} 
will be supposed to hold.


\section{Antecedents}

The first paper describing a defocused two-exposure method to measure out-of-plane rotations was due to Tiziani,\cite{tiziani1} and was later extended for vibration analysis.\cite{tiziani2} If normal illumination and observation are used, the speckle shift at the recording plane is given by\cite{rastogi}
\begin{equation}
d_x= 2 f \beta
\end{equation}
\begin{equation}
d_y= - 2 f \alpha
\end{equation}
where $\alpha,\beta$ are the (small) rotation angles around the $x,y$ axis of a cartesian system placed on the mean plane of the object surface and $f$ is the focal length of the recording system. Lateral displacements do not appreciably alter these values.

Gregory considered divergent illumination and showed that when the optical system is focused on the plane than contains the image of the point source considering the object surface as a mirror, the speckle shift only depends on out-of-plane tilts.\cite{gregory1,gregory2,gregory3} Chiang and Juang described a method to measure the change in slope by defocused systems.\cite{chiangjuang} A great number of later papers document the use of defocused speckle photography to measure in-plane and out-of-plane rotations and strains.\cite{platedeformation,fotovibration,schwieger,sjodahl} 

Today CCD cameras store the specklegrams taken before and after the mechanical transformation in a digital computer and adequate algorithms reveal the speckle shift distribution with sub-pixel accuracy.\cite{sutton,chen,sjo93,amodio} 

\begin{figure}
\begin{center}
\includegraphics[width=10cm]{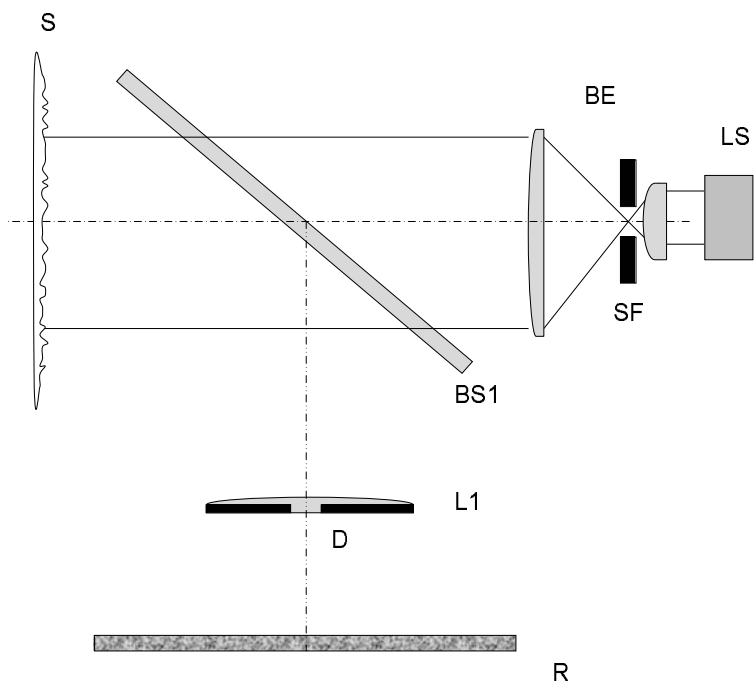}
\end{center}
\caption{DIDDSP set-up.}
\label{fig::diddsp}
\end{figure}

Fig.\ref{fig::diddsp} represents a typical set-up for measuring slope chages in a solid surface. A laser source LS emits a light beam, half of which goes through beam splitter BS and reaches the rough surface S. Part of the light scattered from S is reflected by BS and recorded at the back focal plane R of lens L1. Henceforth, this system will be refered to as DIDDSP(direct illumination digital defocused speckle photography) set-up. 

For a defocused recording system, the speckle size is given by\cite{cloud}
\begin{equation}
s=\frac{1.22 \lambda f}{D}
\end{equation}

When a part of the object surface undergoes an out-of-plane rotation $\alpha$, the light scattered from it experiments a rotation $2\alpha$ and completely falls off the aperture of the system when
\begin{equation}
2\alpha > \frac{D}{d}
\end{equation}
Thus  the maximum measurable rotation is
\begin{equation}
\Gamma = \frac{ D }{2d  }
\end{equation}

Neglecting diffraction effects, the diameter of the area which reflects light to the same point at the detector plane is $D$. Consequently, the lateral resolution $\Delta$ of the measurements is equal to $D$

The relation
\begin{equation}\label{sameh}
\frac{\Gamma}{\Delta} = \frac{1}{2d}
\end{equation}
represents the mutual limitation on range and spatial frequency, that holds for any aperture.

With regard to the requirements posed by the use of a digital system, it follows that if the optimum speckle size is $s^*$, the diameter $D$ should be chosen according to 
\begin{equation} 
D = \frac{1.22 \lambda f}{s^*}
\end{equation} 
and hence,
\begin{equation}
\Gamma = \frac{1.22 f \lambda}{2d s^*}
\end{equation}
\begin{equation}
\Delta = \frac{1.22  \lambda f}{s^*}
\end{equation}


\section{AADDSP}

\begin{figure}
\begin{center}
\includegraphics[width=10cm]{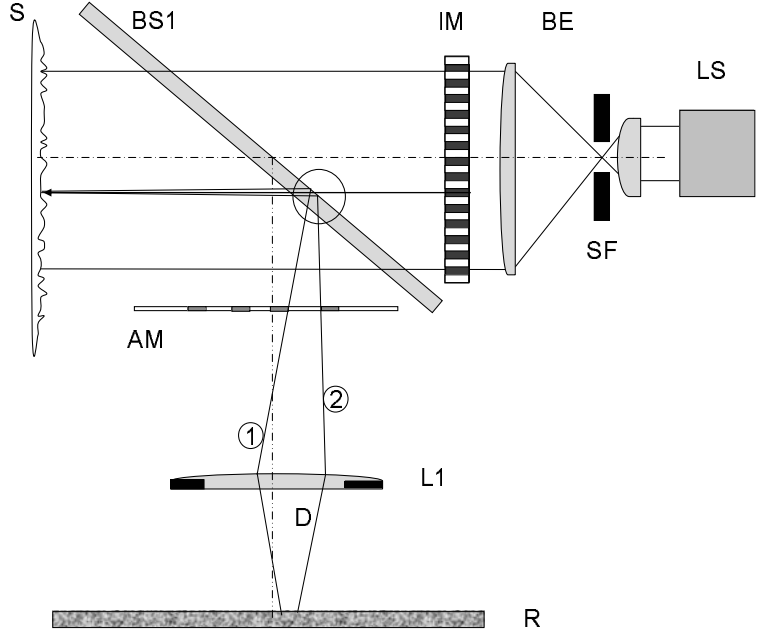}
\end{center}
\caption{AADDSP set-up.}
\label{fig::aaddsp}
\end{figure}

\begin{figure}
\begin{center}
\includegraphics[width=10cm]{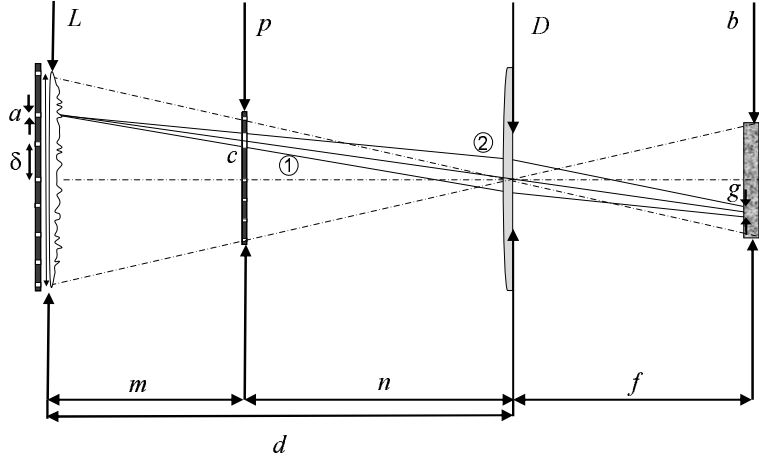}
\end{center}
\caption{AADDSP uniaxial equivalent.}
\label{fig::uniaxial}
\end{figure}

In Fig.\ref{fig::aaddsp} the new system is depicted. A beam from the laser source LS is expanded and spatially filtered to obtain a collimated beam at least as wide as the area under study. The beam is splitted by the perforated mask IM into a set of narrow beams of diameter $a_i$, provided that the wavelength $\lambda$ is much smaller than $a_i$ ($\lambda \ll a_i$). A beam splitter BS lets half of the radiation arrive at the diffuse surface $\Sigma$. The light reflected by $\Sigma$ reaches BS again, half of its intensity goes through a second perforated mask AM and is finally recorded on the back focal plane of lens L1. The circular holes in AM are aligned with the rays coming from the centers of the holes in IM to the center of L1. We will refer to this set-up as Adaptive Aperture Defocused Digital Speckle Photographic (AADDSP) system. An equivalent uniaxial system is depicted in Fig.\ref{fig::uniaxial}.

\begin{figure}
\begin{center}
\includegraphics[width=10cm]{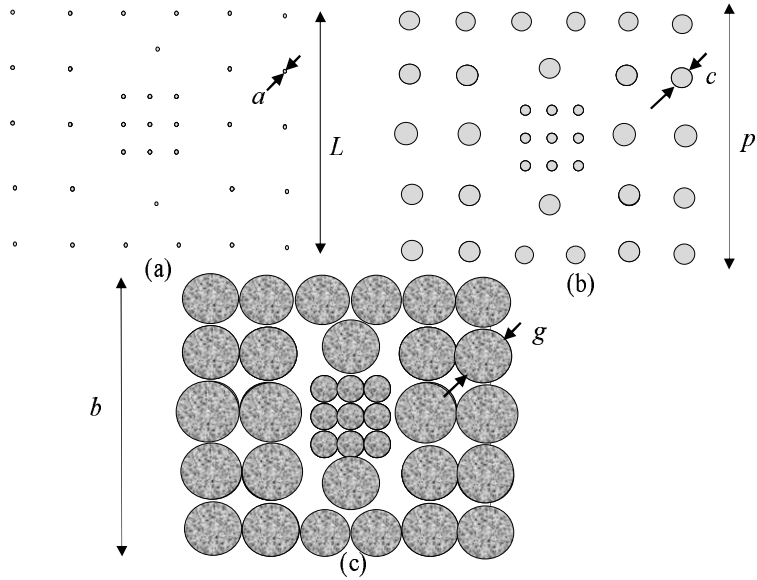}
\end{center}
\caption{a. Illumination mask (IM); b. Aperture mask (AM); c. Speckle circles at the recording plane.}
\label{fig::circulos}
\end{figure}

The Fig.\ref{fig::circulos} represents an IM (a), a corresponding AM (b) and the resulting specklegram (c).

Each collimated beam emerging from IM illuminates a circle $\Sigma_i$ of diameter $a_i$ in the object surface $\Sigma$. From each point in $\Sigma_i$ parts a light cone limited by a corresponding circle $e_i$ of diameter $c_i$ in AM. The aperture diameter $D$ of L1 is to be computed so as to exclude the light that pasess trough any other circle ($e_j, j\neq i$) in AM, yet letting all the radiation from the corresponding circle ($e_i$) reach the recording plane, where it forms a speckled circle $r_i$ of diameter $g_i$.

The speckle size at $r_i$ depends mainly on the narrowest aperture encountered by the cone, so that by assuring the condition
\begin{equation}\label{c4a}
c_i\geq 4a_i
\end{equation} 
the speckle size is made practically independent from $c_i$ and can be approximated by
\begin{equation} 
s_i= \frac{1.22 \lambda f}{a_i}
\end{equation} 
Condition \ref{c4a} also limits the decorrelation and intensity fading that occurs towards the border of the circle. 

The optimum speckle size $s^*$ depends mainly on the camera resolution, so that the diameters $a_i$ should all be equal to
\begin{equation}\label{ess}
a = \frac{1.22 \lambda f}{s^*}
\end{equation} 

Next, it has to be assured that the circles $e_i$ on AM do not overlap, provided that the circles $r_i$ in the recording plane do not. This means that the AM circles diameter $c_i$ and the AM width $p$ must satisfy
\begin{equation} 
\frac p {c_i} \ge \frac b {g_i}
\end{equation} 
where $g_i$ is the diameter of the speckle circle in the detector which is given by
\begin{equation} 
g_i=f \frac {c_i} m
\end{equation} 
and
\begin{equation} 
p=n\frac b f
\end{equation} 
being $m,n$ the distances from IM to the object and the lens surfaces respectively, so that
\begin{equation}\label{cmn}
m\leq n
\end{equation} 
On the other hand, spatial consiterations limit the posible values for $m$
\begin{equation} 
m\geq L
\end{equation} 
so that
\begin{equation} 
L\leq m\leq \frac d 2 = \frac{fL}{2b}
\end{equation} 
which entails
\begin{equation} 
f\geq 2b
\end{equation} 
that is easily satisfied by current popular CCD cameras.

As yet another restriction, it is necessary to avoid laser from an illuminated spot to get through the aperture of another spot and reach the entrance of the system. If the separation between the centers of two neighboring holes at IM is $\delta$, the distance between the centers of the corresponding circles at AM is
\begin{equation} 
\delta_c = \frac{n\delta}{d}
\end{equation} 
that determines an angle for the deviated ray from the chief one
\begin{equation} 
\frac{\delta_c - \displaystyle \frac c 2}{ m }
\end{equation} 
which has to be stopped, thus
\begin{equation} 
\frac D 2 \le d \frac{\delta_c - \displaystyle \frac c 2}{ m } = \frac n m \delta - \frac d {2m} c
\end{equation} 
Nevertheless, the aperture must let laser from the corresponding circles at IM go through AM, so that
\begin{equation} 
D\ge c \frac {d}{m} 
\end{equation} 
Taking into account that the measuring range $\gamma$ for the tilts is given by
\begin{equation}\label{egac}
\gamma = \frac c {2m}
\end{equation} 
we arrive at
\begin{equation} 
\gamma d \leq \frac D 2 \leq \frac n m \delta - \gamma d
\end{equation} 
so that the maximum measuring range for all the surface is 
\begin{equation} 
\Gamma = \frac {D}{2d}
\end{equation} 
which implies a minimum lateral resolution $\Delta$ at the points where the maximum range is allowed, so that
\begin{equation} 
\Delta\geq \frac{2md\Gamma}{n}
\end{equation} 

which is further limited by the non-overlapping condition for the circles at the detector plane
\begin{equation} 
\Delta\geq 2d\Gamma
\end{equation} 
which, taking into account condition \ref{cmn}, is more restrictive, so that 
\begin{equation}\label{egd}
\Delta=2d\Gamma
\end{equation} 
 In order to make condition \ref{c4a} easier to fulfill, and considering Eq.\ref{egac}, the set-up will be arranged so that
 \begin{equation} 
 m=n = \frac d 2
 \end{equation} 
 Substitution for $D$ in Eq.\ref{egd} yields
 \begin{equation}\label{same}
\frac{L}{\Delta} = \frac{b}{2f\Gamma} 
 \end{equation} 
 
 The term $b/f$ in popular cameras is on the order of 0.2. If a maximum measuring range of $1 \times 10^{-2}$ rad is desired then a $10 \times 10$ (or bigger) matrix can be obtained.

The condition \ref{c4a} translates into
\begin{equation} 
4a < 2 \gamma m 
\end{equation} 
which for $\gamma =1 \times 10^{-2}$ rad and $d = 100$ cm yields
\begin{equation} 
a < 2.5 mm
\end{equation} 
which is satisfied for typical detector cells whose width $w$ is on the order of $10$ microns and require values of $a$
\begin{equation} 
a \approx \frac{1.22 f \lambda}{2 w} \approx 0.5 mm
\end{equation} 
assuming $\lambda \approx 0.5$ microns, $f = 20$ mm, $w= 12.2$ microns. Or, from another perspective
\begin{equation}\label{r1}
\gamma > \frac{2.44 \lambda f}{wd}
\end{equation} 
which, after substitution for the previous values, yields
\begin{equation} 
\gamma > 0.0025
\end{equation} 

\section{Discussion}

\begin{figure}
\begin{center}
\includegraphics[width=10cm]{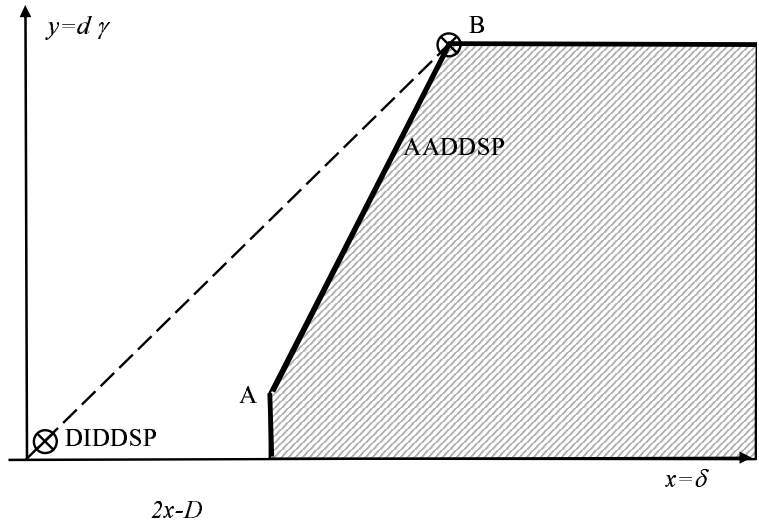}
\end{center}
\caption{Measuring range versus lateral resolution for DIDDSP and AADDSP.}
\label{fig::comp}
\end{figure}

As stated earlier, $\Gamma$ is the maximum measuring range and $\Delta$ is its associated minimum lateral resolution, so that the sampled points in the surface may have better lateral resolutions ($\delta\leq \Delta$), although at the cost of lower ranges ($\gamma \leq \Gamma$). Taking into account Eq.\ref{same} and Eq.\ref{eblfd}, it follows that the possible values of $\Delta,\Gamma$ for AADDSP are mutually limited by Eq.\ref{sameh}, exactly as for DIDDSP. Once $\Gamma,\Delta$ are set in AADDSP, the measuring range $\gamma$ and the lateral resolution $\delta$ for each sampled point in the surface can be chosen taking into account their mutually limiting relation
\begin{equation} 
\gamma \leq \frac \delta d -\Gamma 
\end{equation} 
so that all possible pairs $(\delta,\gamma)$ are those contained in the hatched area in Fig.\ref{fig::comp}. The maximum resolvable spatial frequency corresponds to a lateral resolution $\frac \Delta 2$, with the further restriction arising from condition \ref{r1}. The possible values for $\Gamma_{DI},\Delta_{DI}$ using conventional DIDDSP are those placed on the dashed line in Fig.\ref{fig::comp}. The main advantages of AADDSP are the possibility of having different measuring characteristics throughout the surface and the tunning of $D$ with attention only to the measuring characteristics, because the speckle size is independently set by the choice of $a$.

For $\lambda \approx 0.5$ microns, $f = 20$ mm, $w= 12.2$ microns, $d = 100$ cm, DIDDSP would require values of $D=0.5$ mm and $\Gamma_{DI} = 0.5 \times 10^{-3}$ radians, $\Delta_{DI} = 0.5$ mm. With AADDSP, there is an ample choice for $\Gamma$. If $\Gamma = 1 \times 10^{-2}$, then a minimum resolution $\delta$ can vary from 5 to 10 mm.

Assuming a maximum measuring range $\Gamma$, the value of $D$ is established by the equation
\begin{equation} 
D = 2d\Gamma
\end{equation} 
Accordingly, the possible values of $\delta,\gamma$ lie in the segment joining the points $(\Delta,\Gamma)$, $(\Delta/2, 0)$ excluding those which do not satisfay condition \ref{r1}. This condition may be rewritten as
\begin{equation} 
\gamma \ge 8 \Gamma_{DI}
\end{equation} 
which entails 
\begin{equation} 
\Delta\ge 8 \Delta_{DI}
\end{equation} 
so that for the same optical and recording system, the worst lateral resolution in AADDSP is allways at least eight times poorer than in DIDDSP. However, this drawback may be outweighed by the advantages that will be mentioned in the following. 

As a first advantage of the new method stands its greater range of measurement, which is at least eight times larger than in DIDDSP, assuming the same digital camera. A second advantage is the possibility of different sensitivities in measuring tilts at different points of the surface. The DIDDSP method has one single value for measuring range and lateral resolution, which can be matched by AADDSP in the maximum range points, although this technique allows finer lateral resolutios in other points at the cost of smaller ranges, according to the relation represented by segment AB in Fig.\ref{fig::comp}. 

The third advantage of AADDSP is that the requirements imposed by the recording system (speckle size) can be met without restraining the choice possibilities for range or lateral resolution. This is because the speckle size is set by properly choosing the diameter $a$ of the holes in the illumination mask. In DIDDSP, it was the  speckle size what set the aperture and thus the measuring characteristics. Moreover, in many cases (such as plate bending analysis), the practical values of slope change allowed by the size of the recording cells in DIDDSP fall too short and AADDSP finds its primary applications.

If abstraction is made of speckle and pixel sizes, (for instance, by considering different optic elements) the DIDDSP technique may be tuned so that $\Delta_{DI},\Gamma_{DI}$ may lay on any point of segment AB in Fig.\ref{fig::comp}. For AADDSP, it is a segment (not just a point) what can be chosen, and it is any segment whose slope is twice the one of AB and whose right-top end lies on AB.

 \end{document}